\documentclass[aps,pre,twocolumn]{revtex4}
\usepackage[dvips]{graphicx}
\usepackage{hhline}
\newfont{\myBbb}{msbm10 scaled 1200}
\newcommand{\mod}[1]{\ (\bmod\ #1)}
\newcommand{\Z}{{\mbox{\myBbb Z}}}

\begin{document}

\title{Applying dissipative dynamical systems to pseudorandom number
generation: Equidistribution property and statistical independence
of bits at distances up to logarithm of mesh size}

\author{L.Yu. Barash}

\affiliation{
Landau Institute for Theoretical Physics, 142432 Chernogolovka, Russia\\
e-mail: \tt barash@itp.ac.ru}

\begin{abstract}
The behavior of a family of dissipative dynamical systems representing transformations
of two-dimensional torus is studied on a discrete lattice and compared with that of
conservative hyperbolic automorphisms of the torus. Applying dissipative dynamical systems
to generation of pseudorandom numbers is shown to be advantageous and equidistribution of
probabilities for the sequences of bits can be achieved. A new algorithm for generating
uniform pseudorandom numbers is proposed. The theory of the generator, which includes
proofs of periodic properties and of statistical independence of bits at distances
up to logarithm of mesh size, is presented. Extensive statistical testing using available
test packages demonstrates excellent results, while the speed of the generator is comparable
to other modern generators.
\end{abstract}

\maketitle

\section{Introduction}

Pseudorandom number generation is an important component of
any stochastic simulations
such as molecular dynamics and Monte Carlo simulations~\cite{cm}. The problem
of design of reliable and fast generators is of great importance
and attracts much attention~\cite{Knuth}.

There are numerous papers where chaos is considered as a requirement 
for good pseudorandomness. Many properties of chaotic dynamical systems
are discussed in this respect: ergodicity, sensitivity to initial conditions,
mixing property, local divergence of trajectories, deterministic dynamics 
and structural complexity. These properties resemble certain properties 
of pseudorandomness and are considered in the literature as desirable 
properties for pseudorandomness.
Several pseudorandom number generators based on chaotic maps have been proposed 
in the literature~\cite{chaoticRNGs,BarashShchur}.
However, the behavior of dynamical systems
on a discrete lattice is studied much less than in continuous space
and a number of corresponding important questions
still remain open. In this work I show that applying dissipative dynamical systems 
to pseudorandom number generation can result in substantially preferable statistical 
behavior of the corresponding pseudorandom number sequences, compared to applying 
conservative dynamical systems.

The present approach extends the method of pseudorandom
number generation of Ref.~\cite{BarashShchur,BarashShchur2}, which is
based on evolution of the ensemble of dynamical
systems. Several generalizations are carried out.
The connection between the statistical
properties of a generator and geometric properties of the
corresponding map is uncovered. New pseudorandom number 
generator is proposed. Using SSE2 technology, which is supported
by all Intel and AMD processors fabricated later than in 2003~\cite{Pentium4,AMD64},
effective implementations are developed.

One of the most important properties 
characterizing the quality of pseudorandom sequences of numbers
is the high-dimensional uniformity and the corresponding
equidistribution property~\cite{Equidistr}.
Unlike other essential characteristics of pseudorandom
number generators such as the period length, which is studied
in detail in relationship to nearly all known generators,
there are not so many examples in which the high-dimensional
equidistribution property was proved~\cite{Equidistr,MT,Panneton,LFSR113}.

In this paper the proper choice of parameters is established, 
which results in the validity of the equidistribution property for the proposed generator. 
In particular, it is shown that the determinant of the transformation has to be an even 
integer in order for the property to hold.
The equidistribution is established
on length up to a characteristic length $\ell$:
for $n\leq \ell$,
each combination of successive $n$ bits taken
from the RNG output occurs exactly the same number of times
and has a corresponding probability $1/2^n$.
The length $\ell$ turns out to depend linearly on $t$,
where the mesh size $g$ (i.e. the modulus of the basic recurrence)
is equal to $p\cdot 2^t$ and $p$ is an odd prime.
In other words, for given $p$, one has $\ell\propto\log g$.
Numerical results show that
the equidistribution property
still approximately holds with high accuracy beyond the region of its strict
validity under the condition $n<6.8\log p$.

I have constructed several realizations for the proposed generator~(see Table~\ref{RNGParams}).
It is shown in Proposition 2 in the section on geometric 
and statistical properties that for the realizations
either $\ell=2t-1$ or $\ell=(t-1)/2$ takes place.
The speed and statistical properties of the constructed 
generators are compared with those of other modern generators 
(see Tables~\ref{RNG-Properties}, \ref{Tspeed1}).
Practically, the generators with smaller values of $t$
(e.g. with prime $g$)
also have very good properties for a particular choice of parameters, while
the generator period is not less than $p^2-1$ and increases significantly 
with increasing $p$.
For this reason two realizations with small $t$ are also thoroughly tested.

Among several statistical test suites available in the literature, TestU01 is known 
to contain very stringent batteries of tests for empirical testing of pseudorandom numbers.
At present there are only several known pseudorandom number generators that pass all the tests even 
in the sense that no p-value is outside the interval $[10^{-10},1-10^{-10}]$~\cite{TestU01Paper}.
Statistical testing with TestU01 confirms excellent statistical properties 
of the proposed realizations.

The results obtained have further perspectives in view of 
generating large number of guaranteed statistically independent 
pseudorandom streams, which can be particularly well-suited 
for use in a parallel, distributed environment.

\section{The generator, its initialization and period}
\label{generatorSec}

It is suggested in~\cite{BarashShchur,BarashShchur2} to construct
RNGs based on an ensemble of sequences generated by multiple recursive method.
The state of the generator consists of 
the values $x_i^{(n-1)}, x_i^{(n-2)}\in\{0,1,\dots,g-1\}$, $i=0,1,\dots,s-1$.
The transition function of the generator 
is defined by the recurrence relation
\begin{eqnarray}
x_i^{(n)}&=&kx_i^{(n-1)}-qx_i^{(n-2)}\mod{g},
\label{Recurrence}
\end{eqnarray}
where $i=0,1,\dots,s-1$. 
The values $x_i^{(n)}$, $i=0,1,\dots,s-1$ can be considered as $x$-coordinates 
of $s$ points $(x_i^{(n)},y_i^{(n)})^T$, $i=0,1,\dots,s-1$ of the $g\times g$ lattice 
on the two-dimensional torus, then each recurrence relation 
describes the dynamics of $x$-coordinate of a point 
on the two-dimensional torus:
\begin{equation}
{x_i^{(n)}\choose y_i^{(n)}}=M
{x_i^{(n-1)}\choose y_i^{(n-1)}}\mod{g},
\label{MatrixRecurr}
\end{equation}
where matrix $M={{m_1\ m_2\choose m_3\ m_4}}$
is a matrix with integer elements, $k={\rm{Tr}}\, M$,
$q=\det M$ and ${\rm{Tr}}\, M$ is a trace of matrix $M$~\cite{Grothe,Lecuyer90,BarashShchur}.
Indeed, it follows from (\ref{MatrixRecurr}) that
$kx_i^{(n-1)}-qx_i^{(n-2)}=(m_1+m_4)x_i^{(n-1)}-(m_1m_4-m_2m_3)x_i^{(n-2)}=
(x_i^{(n)}-m_2y_i^{(n-1)})+m_4x_i^{(n-1)}-m_1m_4x_i^{(n-2)}+m_2m_3x_i^{(n-2)}=
x_i^{(n)}-m_2(y_i^{(n-1)}-m_3x_i^{(n-2)})+m_4(x_i^{(n-1)}-m_1x_i^{(n-2)})=
x_i^{(n)}-m_2m_4y_i^{(n-2)}+m_2m_4y_i^{(n-2)}=x_i^{(n)}\mod{g}$.
The basic recurrence (\ref{Recurrence}) is therefore closely related
to so-called matrix generator of pseudorandom numbers
studied in~\cite{Knuth,Grothe,NiederrBook}.

The output function is defined as follows:
\begin{equation}
a^{(n)}=\sum_{i=0}^{s-1} \lfloor 2x_i^{(n)}/g\rfloor \cdot 2^i,
\label{OutputFunction}
\end{equation}
where $i=0,1,\dots,s-1$, i.e. each bit
of the output corresponds to its own recurrence,
and $s=32$ recurrences are calculated in parallel.

For $g=p\cdot 2^t$, where $p$ is a prime number,
the characteristic polynomial $f(x)=x^2-kx+q$ is chosen to be
primitive over $\Z_p$.
Primitivity of the characteristic polynomial guarantees maximal
possible period $p^2-1$ of the output sequence for $g=p$.
It is straightforward to prove that taking $g=p\cdot 2^t$ 
instead of $g=p$ does not reduce the value of the period.

There is an easy algorithm to calculate $x^{(n)}$ in (\ref{Recurrence})
very quickly from $x^{(0)}$ and $x^{(1)}$ for any large $n$.
Indeed, if $x^{(2n)}=k_n x^{(n)}-q_n x^{(0)}\mod{g}$, then
$x^{(4n)}=(k_n^2-2q_n) x^{(2n)}-q_n^2 x^{(0)}\mod{g}$.
As was mentioned already in~\cite{BarashShchur}, this helps to
initialize the generator. To initialize~all $s$ recurrences,
the following initial conditions are used:
$x_i^{(0)}=x^{(iA)}$, $x_i^{(1)}=x^{(iA+1)}, i=0,1,\dots,s-1$.
Here $A$ is a value of the order of $(p^2{-}1)/s$.
The author has tested realizations with various values
of $A$ of the order of $(p^2{-}1)/s$ and found
in all cases that the specific choice of $A$ was not
of importance for the properties studied in the next sections.
Short cycles and, in particular, the cycle consisting of zeroes,
are avoided if at least one of $x^{(0)}$ and $x^{(1)}$ is
not divisible by $p$. As a result of the initialization,
all $s$ initial points belong to the same orbit on the torus
of the period $p^2-1$, while the minimal distance $A$
between the initial points along the orbit is chosen to be very large.

\begin{table}[t]
\caption{Parameters of the new generators.}
\begin{center}
\small
\begin{tabular}{|l|c|c|c|c|c|c|}
\hline
Generator & $g$ & $k$ & $q$ & $v$ & Period \\
\hline
GM29.1-SSE  & $2^{29}-3        $ & $4  $ & $2  $ & $1$ & $= 2.8\cdot 10^{17}   $ \\
GM55.4-SSE  & $16(2^{51}-129)  $ & \hspace{-1mm}$256$\hspace{-1mm} & \hspace{-1mm}$176$\hspace{-1mm} & $4$ & $\geq 5.1\cdot 10^{30}$ \\
GQ58.1-SSE  & $2^{29}(2^{29}-3)$ & $8  $ & $48 $ & $1$ & $\geq 2.8\cdot 10^{17}$ \\
GQ58.3-SSE  & $2^{29}(2^{29}-3)$ & $8  $ & $48 $ & $3$ & $\geq 2.8\cdot 10^{17}$ \\
GQ58.4-SSE  & $2^{29}(2^{29}-3)$ & $8  $ & $48 $ & $4$ & $\geq 2.8\cdot 10^{17}$ \\
\hline
\end{tabular}
\end{center}
\label{RNGParams}
\end{table}

In addition to the realizations based on the output function~(\ref{OutputFunction})
that takes a single bit from each linear recurrence, I have also constructed
realizations based on a more general output function
\begin{equation}
a^{(n)}=\sum_{i=0}^{s-1} \lfloor 2^v x_i^{(n)}/g\rfloor \cdot 2^{iv},
\label{OutputFunction2}
\end{equation}
where $v$ bits are taken from each recurrence and $i=0,1,\dots,s-1$. 
For example, GM55.4-SSE realization 
calculates only $s=8$ recurrent relations in parallel 
and takes $v=4$ bits from each number.
Pseudorandom 32-bit numbers can be generated if $sv \geq 32$.
The sequence of bits $\{\lfloor 2^vx_i^{(n)}/g\rfloor\}$,
where $i$ is fixed and $\{x_i^{(n)}\}$ is generated with 
relation (\ref{MatrixRecurr}) will be designated below
as a stream of $v$-bit blocks generated with matrix $M$.
The pairs $x_i^{(0)}, x_i^{(1)} \in \Z_g$ for the recurrence (\ref{Recurrence}) 
and $x_i^{(0)},y_i^{(0)}\in \Z_g$ for the recurrence (\ref{MatrixRecurr})
represent seeds for the streams of $v$-bit blocks generated with (\ref{Recurrence}) 
and (\ref{MatrixRecurr}) respectively.
Consider the set of admissible seeds containing all seeds such that 
at least one of the two values is not divisible by $p$.
Selecting the seed at random from a uniform distribution
over the set of admissible seeds determines the probability measure 
for output subsequences of a stream of $v$-bit blocks.
Such probabilities are considered below in the next section.

The parameters for the particular constructed realizations 
of the generator are shown in Table~\ref{RNGParams}.
The parameters are chosen in order for the characteristic polynomial
$x^2-kx+q$ to be primitive over $\Z_p$. 
In addition, as is shown below, value of $q$ 
must be divisible by $2^v$ in order for the equidistribution property to hold.
Also the value of $(k+q)g$ should not exceed either $2^{32}$ or $2^{64}$
in order to effectively calculate four 32-bit recurrences 
or two 64-bit recurrences in parallel within SIMD arithmetic.
In the particular case $t=0$ and $v=1$ the method reduces
to that studied earlier in~\cite{BarashShchur,BarashShchur2}.
Program codes for the new generators and proper initializations
are available in~\cite{AlgSite}.

\section{Geometric properties and statistical properties}
\label{geomANDstat}

In~\cite{BarashShchur} a connection is established between statistical properties,
the results of a random walk test and geometric properties of the cat maps.
Cat maps are simple chaotic dynamical systems that
correspond to transformations~(\ref{MatrixRecurr})
for $q=\det M=1$, i.e. hyperbolic automorphisms of the two-dimensional torus.
In particular, it is proved in~\cite{BarashShchur} that the probability
of sequence $0000$ of the first bits generated by a single cat map
depends only on the trace $k$ of a matrix $M$ and for even $k$
is equal to $P=P_0 k^2/(k^2-1)$, 
where $P_0=1/16$.
If $k$ is odd, then all sequences of length $4$ are equiprobable.
The probability of sequence $00000$ of length $5$ is equal to
$P=P_0(1+1/(3k^2-6))$ for odd $k$, where $P_0=1/32$.
The condition $P>P_0$ signifies that the 5-dimensional equidistribution
never takes place for $q=1$, i.e. for conservative hyperbolic automorphisms of the torus.
In this work a more general case $q\ne 1$ involving dissipative dynamical systems is studied.

\begin{figure}[t]
\caption{(Color online) The regions on the torus obtained in~\cite{BarashShchur}
for the third points of sequences of length 5 for the matrix ${1\ 1\choose1\ 2}$. 
Coordinates $x/g,y/g$ are used.
These regions correspond to the sequences
of length 5 of the first bits generated by the corresponding RNG. 
Each region is drawn with its own color.}
\centering
\includegraphics[width=0.34\textwidth]{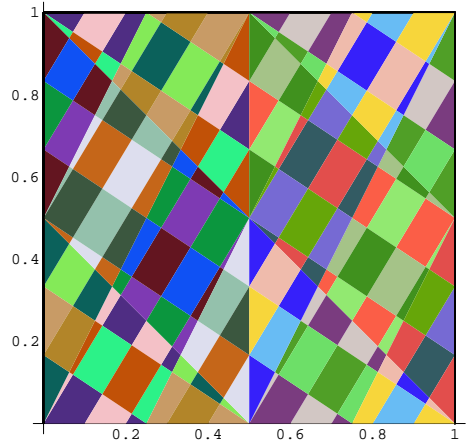}
\label{length5}
\end{figure}

Fig.~\ref{length5} shows the regions on the torus obtained in~\cite{BarashShchur}
for the third points of sequences of length 5 for the matrix ${1\ 1\choose1\ 2}$. 
The regions correspond to the sequences of length 5 of the first bits 
generated by the respective RNG, and the areas of the regions are equal to
the probabilities of the sequences. Each region is drawn with its own color.

Let $X_i=\{ (x,y)^T | i/2^v\leq x/g< (i+1)/2^v, 0\leq y/g <1  \}$, i.e.
the torus is divided into $2^v$ vertical stripes $X_0, X_1, \dots, X_{2^v-1}$ 
of equal area. Suppose that $g$ is divisible by $2^v$ and 
consider the shift $S: (x,y)^T \rightarrow (x+g/2^v,y)^T \mod{g}$,
i.e. $S(X_i)=X_{(i+1)\mod{2^v}}$. The shift $S$ is a superposition
of two rotations: $S=R_2R_1$, where
$R_1$ is a 180-degree rotation with respect to the point $(1/2^{v+1},1/2)^T$
and $R_2$ a 180-degree rotation with respect to the point $(1/2^v,1/2)^T$.

{\bf Proposition 1.} If
(i) $M={{m_1\ m_2\choose m_3\ m_4}}$ is a matrix
with integer values $m_1$, $m_2$, $m_3$, $m_4$,
(ii) $m_1$, $q=\det M$ and $g$ are divisible by $2^v$,
(iii) the image of the lattice $g\times g$
with the transformation $M^j$ is invariant with respect to
the shift $S$ for $j=0,1,\dots,n$,
then all the sequences of length $n$ in a stream of $v$-bit blocks
generated with matrix $M$ are equiprobable.

{\bf Proof.}
In this case the element $m_1^{(n)}$ of matrix 
\begin{equation}
M^n={m_1^{(n)}\ m_2^{(n)}\choose m_3^{(n)}\ m_4^{(n)}} \mod{g}
\label{Mn}
\end{equation}
satisfies the recurrence relation
$m_1^{(n)}=km_1^{(n-1)}-qm_1^{(n-2)}\mod{g}$. 
Hence $m_1^{(n)}$ is divisible by $2^v$ for any integer $n\geq 1$.

Since $m_1^{(n)}$ is divisible by $2^v$, one has
$M^nS(x,y)^T=M^n(x+g/2^v\mod{g},y)^T=M^n(x,y)^T+(0,m_3^{(n)}g/2^v)^T$. Hence,
the set of points $A$ such that $A\in X_i$ and $M^n(A)\in X_j$
passes with the shift $S$ into the set of points $A$ such that $A\in X_{(i+1)\mod{2^v}}$
and $M^n(A)\in X_j$.

Let's now prove by induction that all sequences of length $n$ are equiprobable.
Obviously, if $g$ is divisible by $2^v$, sequences of length $1$ are equiprobable: 
$P(0)=P(1)=\dots=P(2^v-1)=1/2^v$.
Assume that all sequences of length $n-1$ are equiprobable.
Let $\alpha_i=P({ix_1\dots x_{n-1}})$, $i=0,1,\dots,2^v-1$ be probabilities of
sequences of length $n$. Then $\alpha_i=\alpha_{i+1}$, $i=0,1,\dots,2^v-2$ because
the set of points $A$ of the lattice $g\times g$ such that
$A\in X_i$, $M(A)\in X_{x_1}$,\dots, $M^{n-1}(A)\in X_{x_{n-1}}$ passes with the shift $S$
into the set of points $A$ of the lattice $g\times g$ such that
$A\in X_{(i+1)\mod{2^v}}$, $M(A)\in X_{x_1}$,\dots, $M^{n-1}(A)\in X_{x_{n-1}}$.
On the other hand, $\sum_{i=0}^{2^v-1}\alpha_i$
is the probability of sequence $x_1\dots x_{n-1}$ of length $n-1$ and equals $1/2^{v(n-1)}$.
Therefore, $\alpha_i=1/2^{vn}, i=0,1,\dots,2^v-1$, and all the sequences of length $n$
are equiprobable. Proposition 1 is proved.

\begin{figure*}[t]
\caption{ The set of points $A$ such that $A\in X_0$ and $M^2(A)\in X_0$ (left panel) and 
the set of points $A$ such that $A\in X_1$ and $M^2(A)\in X_0$ (right panel)
for $M={2\ 2\choose1\ 2}$ and $v=1$. Coordinates $x/g,y/g$ are used.}
\centering
\includegraphics[width=0.35\textwidth]{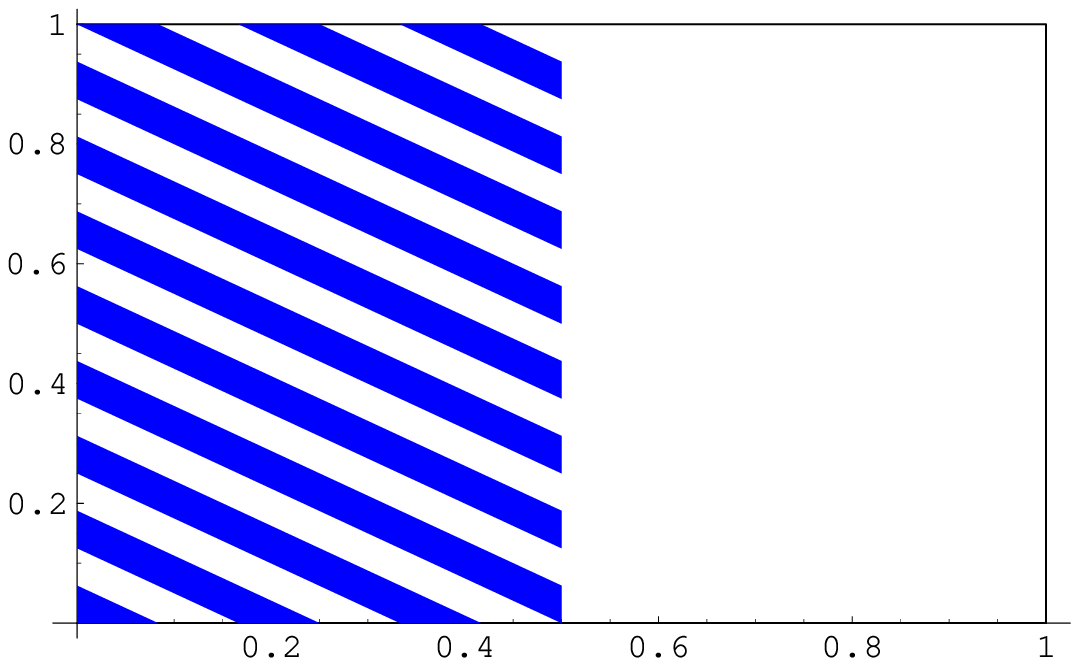}
\includegraphics[width=0.35\textwidth]{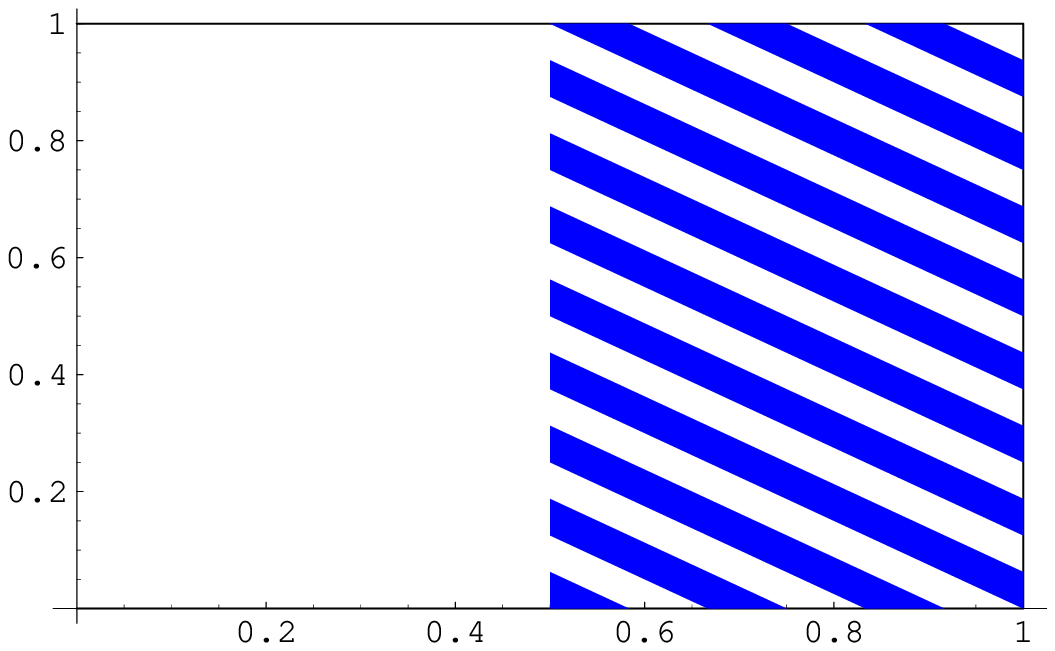}
\label{regions}
\end{figure*}

The condition that the image of the lattice $g\times g$ with the transformation
$M^j$ is invariant with respect to the shift $S$
for $j=0,1,\dots,n$, is used in the above consideration and is necessary
for the Proposition 1. For $j=0$ the invariance means that $g$ is divisible by $2^v$.
If $g$ and $m_1^{(n)}$ are divisible by $2^v$, then
the number of points $A$ of the lattice $g\times g$
such that $A\in X_0$ and $M^n(A)\in X_0$ is equal to the number of points $A$ 
of the same lattice such that $A\in X_1$ and $M^n(A)\in X_0$. If $g$ is not divisible
by $2^v$ then these numbers are approximately equal because the corresponding
areas are equal and $g$ is large number, 
and the exact equality holds only if $g$ is divisible by $2^v$.
Fig.~\ref{regions} shows the sets of points 
$\{A|A\in X_0, M^2(A)\in X_0\}$ and $\{A|A\in X_1, M^2(A)\in X_0\}$
for $M={{2\ 2}\choose {1\ 2}}$ and $v=1$.

{\bf Proposition 2.}
For $M={{2\ 2}\choose{1\ 2}}$, 
$M={{10\ 17}\choose{-4\ -2}}$
and $M={{244\ 43}\choose{32\ 12}}$
the sequences of length $1,2,\dots,\ell$
in a stream of bits generated with matrix $M$
are equiprobable, where $\ell=2t-1$,
$\ell=(t-1)/2$ and $\ell=(t-1)/2$ respectively.
Here $g=p\cdot 2^t$, where $p$ is an odd prime,
and the matrices correspond to the realizations 
GM29-SSE, GM58-SSE and GM55-SSE respectively.

{\bf Proof.}
Let's check that
the image of the lattice $g\times g$ with the transformation
$M^j$ is invariant with respect to the shift
for $j=0,1,\dots,n$ and $n\leq \ell$.
In particular, the invariance takes place if
there are integers $r,l<t$ such that
the distance between integer vectors $(x+g/2^{r+1},y+g/2^{l+1})^T$
and $(x,y)^T$ after applying transformation $M^j$ is equal to 
$(g/2,0)^T$ modulo $g$. This results in
$
{(
{m_1^{(j)}}/{2^r}+{m_2^{(j)}}/{2^l}
,
{m_3^{(j)}}/{2^r}+{m_4^{(j)}}/{2^l}
)^T}
\equiv
{(1, 0)^T} \mod{2}
$. For the matrix $M={{2\ 2}\choose{1\ 2}}$ the condition is satisfied when
$r=j/2$, $l=j/2-1$ for even $j$ and $r=(j-1)/2$, $l=(j+1)/2$ for odd $j$.
Thus $\ell=j_{max}+1=2t-1$. Similarly, for each of the matrices
$M={{10\ 17}\choose{-4\ -2}}$ and $M={{244\ 43}\choose{32\ 12}}$
the condition is satisfied for $\ell=(t-1)/2$. Proposition 2 is proved.

Generally, the following statements are also valid.
Consider a matrix $M$ with integer elements and 
the following integer quantities:
$g=p\cdot 2^t$, $q=\det M=2^u w\mod{g}$, 
$k={{\rm{Tr}}\, M}=2^m r\mod{g}$, $u\geq 1$, $t\geq v$, $m\geq 0$.
Here $w,r$ are odd integers and $p$ is an odd prime. Then
(i) all $2^j$ sequences of length $j$ in 
a stream of $v$-bit blocks generated with recurrent relation 
(\ref{Recurrence}) are equiprobable for $j=1,2,\dots,\ell$. Here
$\ell=\lceil (t-v)/\lceil u/2\rceil\rceil$ for $u\leq 2m$
and $\ell=\lceil(t-v)/(u-m)\rceil$ for $u>2m$;
(ii) if $k$ is even, then the image of the lattice $g\times g$
with the transformation $M^{2t}$ is the lattice $p\times p$ on the torus;
(iii) if $k$ is odd, then the image of the lattice $g\times g$
with the transformation $M^{\lceil t/u\rceil}$ is not invariant
with respect to the shift $S$.

Although the exact equidistribution property does not hold 
when distance between some points of the sequence $\geq 2t$, numerical results show
that the equidistribution holds approximately with high accuracy
for the sequences of bits of length $n$, where $n<6.8\log p$.
Also, one can take $n$ points
with arbitrary distances (not exceeding $p^2-1$) between them along 
the orbit (i.e. not necessarily successive points of the orbit), where $n<6.8\log p$,
and still the approximate equidistribution will hold with a high accuracy.
The output value $a^{(n)}$ in (\ref{OutputFunction})
consists of high-order bits of $s=32$ successive points along the orbit
of matrix $M^A$, where $A$ is the value of the order of $(p^2-1)/s$.
Therefore, according to the numerical results, the output value $a^{(n)}$ has
a uniform distribution with a very high accuracy.

\begin{figure}[t]
\begin{center}
\includegraphics[width=0.35\textwidth]{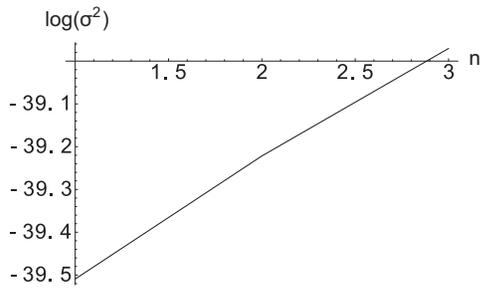}
\end{center}
\caption{Variance of the numbers of points of the $p\times p$-lattice
corresponding to sequences of length $n$ versus $n$. The values 
are normalized such that $\langle A_n\rangle=1$.}
\label{variance}
\end{figure}

In most cases the image of the lattice $g\times g$
on the torus with $M^j$ where $j\geq 2t$ is the $p\times p$-lattice, therefore
it is most interesting to study the deviations from the equidistribution 
for the $p\times p$-lattice.
I have calculated the exact areas on the torus which correspond to each of the
sequences for $M={{1\ 1}\choose{1\ 3}}$. The calculations were carried out
on a PC using Class Library for Numbers~\cite{CLN} for exact rational arithmetics.
For each of the $2^n$ sequences of length $n=1,2,\dots$, 
the corresponding set of points on the unit two-dimensional torus consists of filled polygons.
Exact rational coordinates of all the vertices of each filled polygon
were found. Also, the exact number of points of the $p\times p$
lattice inside each polygon was calculated.
The total area of the polygons for each of the $2^n$ sequences of length $n$
was found to equal $1/(2^n)$. Such equality of the areas for different sequences
of the same length was observed for matrices
with even determinant and was not observed for matrices with odd determinant. Let 
$A_{n,0}, A_{n,1},\dots, A_{n,2^n-1}$ be the numbers
of points of the $p\times p$-lattice corresponding to the sequences of length $n$.
Then $\sum_{i=0}^{2^n-1} A_{n,i}=p^2$. Therefore, if $A_n$ is the set of numbers
$A_n=\{2^n A_{n,0}/p^2, 2^n A_{n,1}/p^2,\dots, 2^n A_{n,2^n-1}/p^2\}$, 
then $\langle A_n\rangle=1$, where $\langle A_n\rangle$
is the average value of $A_n$.
The dependence of logarithm of variance of $A_n$ on $n$ is shown in Fig.~\ref{variance}
for $p=2^{29}-3$. The calculations for smaller values of $p$ and larger values of $n$ 
demonstrate that the dependence of $\log(\sigma^2)$ on $n$ is almost linear.
Calculations show that 
the deviations from equidistribution are negligibly small in the sence that
$\sigma(A_n)$ is much smaller than $\langle A_n\rangle=1$,
for $n<6.8 \log p$. In particular, for $p=2^{29}-3$ the deviations
are small for $n<130$.

The variance for the several points of the orbit
of matrix $M$ on the $p\times p$-lattice on the torus, 
is found to substantially depend on the number of points and on the value of $p$,
and only weakly depend (within several percent) on the distances between 
the points along the orbit.

\section{Statistical testing} 

Table~\ref{RNG-Properties} shows the results of applying the SmallCrush,
PseudoDiehard, Crush and BigCrush batteries of tests taken from~\cite{TestU01},
to the generators introduced in Table~\ref{RNGParams}.
Batteries SmallCrush, PseudoDiehard, Crush and BigCrush contain 15, 126, 144
and 160 statistical tests respectively.
For each battery of tests, Table~\ref{RNG-Properties} displays
three characteristics: the number
of statistical tests with p-values outside the
interval $[10^{-3},1-10^{-3}]$, number of tests with p-values
outside the interval $[10^{-5},1-10^{-5}]$, and number of tests
with p-values outside the interval $[10^{-10},1-10^{-10}]$.
Table~\ref{RNG-Properties} also contains the results of statistical 
tests for Mersenne Twister generator of Matsumoto and Nishimira~\cite{MT},
combined Tausworthe generator of L'Ecuyer~\cite{LFSR113} and
combined multiple recursive generator proposed in~\cite{CombinedLCG}.
These generators are modern examples of fast RNG implementations with good
statistical properties (see Sec. 4.5.4 and Sec. 4.6.1 in~\cite{LEcuyerMain}).
Both LFSR113 and MT19937 fail the test \verb1scomp_LinearComp1
that is a linear complexity
test for the binary sequences~(see \cite{TestU01}),
because the bits of LFSR113 and MT19937
have a linear structure by construction.
Also LFSR113 fails the test \verb1smarsa_MatrixRank1~(see \cite{TestU01}).
The period lengths for the generators MRG32K3A, LFSR113 and MT19937
are $3.1\cdot 10^{57}$, $1.0\cdot 10^{34}$ and $4.3\cdot 10^{6001}$ respectively.

The usefulness of a RNG for a specific application
in physics depends on, possibly dangerous interferences of
the correlations in the specific problem and those of the RNG.
Modern statistical test suites contain tests that reveal 
known types of correlations for the RNGs, in particular,
the types that are known to result in systematic errors in Monte-Carlo
simulations and that were studied in~\cite{MCerrors}.
One concludes that the new realizations described in this paper
possess excellent statistical properties.

\begin{table}[tb]
\caption{Numbers of failed tests
for the batteries of tests SmallCrush, Crush, BigCrush~\cite{TestU01},
and Diehard~\cite{TestU01}.
Testing was performed with
package TestU01 version TestU01-1.2.3.
For each battery of tests, three numbers are displayed: the number
of statistical tests with p-values outside the
interval $[10^{-3},1-10^{-3}]$, number of tests with p-values
outside the interval $[10^{-5},1-10^{-5}]$, and number of tests
with p-values outside the interval $[10^{-10},1-10^{-10}]$.
}
\begin{center}
\small
\begin{tabular}{|l|c|c|c|c|c|c|}
\hline
Generator & SmallCrush & Diehard & Crush & \hspace{-0.03cm}BigCrush\hspace{-0.03cm} \\
\hline
MRG32k3a  & $ 0, 0, 0$ & $ 0, 0, 0$ & $ 0, 0, 0 $ & $ 0, 0, 0$ \\
LFSR113   & $ 0, 0, 0$ & $ 1, 0, 0$ & $ 6, 6, 6 $ & $ 6, 6, 6$ \\
MT19937   & $ 0, 0, 0$ & $ 0, 0, 0$ & $ 2, 2, 2 $ & $ 2, 2, 2$ \\
GM29.1-SSE  & $ 0, 0, 0$ & $ 0, 0, 0$ & $ 0, 0, 0$ & $ 0, 0, 0$ \\
GM55.4-SSE  & $ 0, 0, 0$ & $ 0, 0, 0$ & $ 0, 0, 0$ & $ 0, 0, 0$ \\
GQ58.1-SSE  & $ 0, 0, 0$ & $ 0, 0, 0$ & $ 0, 0, 0$ & $ 0, 0, 0$ \\
GQ58.3-SSE  & $ 0, 0, 0$ & $ 0, 0, 0$ & $ 0, 0, 0$ & $ 0, 0, 0$ \\
GQ58.4-SSE  & $ 0, 0, 0$ & $ 0, 0, 0$ & $ 0, 0, 0$ & $ 0, 0, 0$ \\
\hline
\end{tabular}
\end{center}
\label{RNG-Properties}
\end{table}

\section{Speed of the generator}
\label{SpeedSec}

\begin{table*}[htb]
\caption{CPU time (sec.) for generating $10^9$ random numbers. Processors:
~Intel Core~i7-940 and~AMD Turion X2 RM-70. Compilers: gcc 4.3.3, icc 11.0.}
\vspace{0.2cm}
\label{Tspeed1}
\small
\begin{center}
\begin{tabular}{|l||c|c|c|c||c|c|c|c||c|}
\hline
Intel Core i7-940 & gcc -O0 & gcc -O1 & gcc -O2 & gcc -O3 & icc -O0 & icc -O1 & icc -O2 & icc -O3 & Source \\
\hline
MT19937          &  13.7   &  5.7    &  6.9    &   2.6   &  17.5   &   6.5   &   2.9   &  2.9  & \cite{MT} \\
MT19937-SSE      &   5.2   &  4.8    &  5.5    &   2.0   &   4.9   &   4.7   &   2.4   &  2.0  & \cite{BarashShchur2} \\
LFSR113          &  10.4   &  4.8    &  6.8    &   3.1   &  10.2   &   5.0   &   4.6   &  4.5  & \cite{LFSR113} \\
LFSR113-SSE      &   8.0   &  6.8    &  6.8    &   6.9   &   7.3   &   6.9   &   6.6   &  6.5  & \cite{BarashShchur2} \\
MRG32k3a         &  47.9   &  36.3   &  35.3   &  25.0   &  56.1   &  33.1   &  22.8   &  28.1 & \cite{CombinedLCG} \\
MRG32k3a-SSE     &   9.1   &   7.4   &   5.8   &   5.8   &   8.8   &   7.4   &   6.0   &   5.9 & \cite{BarashShchur2} \\
GM29.1-SSE       &  22.6   &  19.6   &  17.5   &  18.1   &  21.2   &  18.7   &  18.2   & 18.1  & \cite{AlgSite} \\
GM55.4-SSE       &  18.0   &  16.8   &  15.4   &  15.4   &  17.7   &  16.3   &  15.8   & 15.7  & \cite{AlgSite} \\
GQ58.1-SSE       &  50.5   &  49.2   &  47.4   &  47.3   &  50.5   &  48.1   &  48.0   & 47.7  & \cite{AlgSite} \\
GQ58.3-SSE       &  22.0   &  21.2   &  19.0   &  20.1   &  22.5   &  20.4   &  19.5   & 19.5  & \cite{AlgSite} \\
GQ58.4-SSE       &  16.1   &  14.7   &  12.8   &  13.8   &  15.5   &  13.9   &  13.3   & 13.3  & \cite{AlgSite} \\
\hline
\hline
AMD Turion X2 RM-70 & gcc -O0 & gcc -O1 & gcc -O2 & gcc -O3 & icc -O0 & icc -O1 & icc -O2 & icc -O3 & Source \\
\hline
MT19937          &  31.0   &  17.8   &  10.8   &   7.1   &  31.0   &  18.7   &   5.2   &   4.9 & \cite{MT} \\
MT19937-SSE      &  11.3   &  10.3   &  11.1   &   6.6   &  10.8   &   9.9   &   6.0   &   6.0 & \cite{BarashShchur2} \\
LFSR113          &  14.6   &   8.7   &   9.6   &   5.3   &  14.9   &   9.1   &   6.9   &   6.8 & \cite{LFSR113} \\
MRG32k3a         &  89.0   &  60.9   &  60.9   &  47.0   &  89.1   &  69.2   &  41.5   &  41.6 & \cite{CombinedLCG} \\
MRG32k3a-SSE     &  25.9   &  22.3   &  18.4   &  18.3   &  25.6   &  22.3   &  19.0   &  19.0 & \cite{BarashShchur2} \\
GM29.1-SSE       &  68.5   &  64.4   &  60.7   &  60.7   &  67.8   &  63.1   &  61.7   & 61.7  & \cite{AlgSite} \\
GM55.4-SSE       &  59.8   &  54.8   &  53.1   &  53.0   &  58.2   &  53.6   &  52.8   &  52.8 & \cite{AlgSite} \\
GQ58.1-SSE       & 179.6   &  179.6  &  178.3  &  177.8  & 183.1   & 178.3   &  178.5  & 178.5 & \cite{AlgSite} \\
GQ58.3-SSE       &  75.5   &  73.9   &  70.6   &  71.1   &  74.2   &  71.9   &  70.4   & 70.1  & \cite{AlgSite} \\
GQ58.4-SSE       &  51.9   &  51.0   &  48.2   &  48.1   &  53.1   &  49.4   &  48.2   & 48.1  & \cite{AlgSite} \\
\hline
\end{tabular}
\end{center}
\end{table*}

I have tested the CPU times needed for
generating $10^9$ random numbers. The results are shown in Table~\ref{Tspeed1}
for Intel Core i7-940 and AMD Turion X2 RM-70 processors respectively.
The results are presented for different compilers
and optimization options. The compilers in use are GNU C compiler gcc version 4.3.3
and Intel C compiler icc version 11.0.
The CPU times for the realizations 
GM29.1-SSE, GM55.4-SSE, GQ58.1-SSE, GQ58.3-SSE and GQ58.4-SSE
introduced in Table~\ref{RNGParams} are compared with
those for Mersenne Twister generator of Matsumoto and Nishimira~\cite{MT},
combined Tausworthe generator of L'Ecuyer~\cite{LFSR113} and
combined multiple recursive generator proposed in~\cite{CombinedLCG}.

\acknowledgments

The author thanks L.~N.~Shchur for useful discussions and remarks.
The support of RFBR grant 11-07-00197 is acknowledged.


\begin{thebibliography}{99}

\bibitem{cm} K.S.D.\ Beach, P.A.\ Lee, P.\ Monthoux, Phys. Rev.
Lett. {\bf 92} (2004) 026401; D.P.\ Landau and K.\ Binder, {\it A
Guide to Monte Carlo Simulations in Statistical Physics} (Cambridge
University Press, Cambridge, 2000); S.C.\ Pieper and R.B.\ Wiring,
Ann. Rev. Nucl. Part. Sci., {\bf 51} (2001) 53;A.\ L\"uchow, Ann. Rev.
Phys. Chem., {\bf 51} (2000) 501; A.R.\ Bizzarri, J. Phys.: Cond. Mat.
{\bf 16} (2004) R83.

\bibitem{Knuth} D.E.\ Knuth, {\em The Art of Computer Programming},
Vol.\ 2 (Addison-Wesley, Reading, Mass., 3rd edition, 1997).

\bibitem{chaoticRNGs}
M.\ Falcioni et. al., Phys. Rev. E 72, 016220 (2005);
R.\ Lozi, Ind. J. Ind. Appl. Math. 1 (1), 1 (2008);
L.\ Kocarev, G. Jakimoski, IEEE Trans. Circ. Syst. 50(1), 123 (2003);
P.\ Li et. al., Phys. Lett. A 349, 467 (2006);
V.\ Patidar, K.K.\ Sud, EJTP 6 (20), 327 (2009);
N.K.\ Pareek et. al., Int. J. Netw. Sec. 10 (1), 32 (2010).


\bibitem{BarashShchur} 
L.\ Barash, L.N.\ Shchur, Phys. Rev. E {\bf 73} (2006), 036701.

\bibitem{BarashShchur2}
L.Yu. Barash and L.N. Shchur, Comput. Phys. Commun. {\bf 182} (2011), 1518.

\bibitem{Pentium4}
\verb#http://www.intel.com/support/processors/pentium4#
\verb#/sb/CS-029967.htm#

\bibitem{AMD64}
\verb#http://support.amd.com/us/Processor_TechDocs#
\verb#/24592.pdf#

\bibitem{Equidistr} J.P.R.\ Tootil, W.D.\ Robinson, D.J.\ Eagle, J. ACM 20, 3, 469 (1973);
M.\ Fushimi, S. Tezuka, Commun. ACM 4, 516 (1983);
R.\ Couture, P. L'Ecuyer, S. Tezuka, Math. Comput. 60, 749 (1993);
S.\ Tezuka, P. L'Ecuyer, ACM Trans. Model. Comput. Simul. 1, 99 (1991);
P.\ L'Ecuyer, Math. Comput. 65, 203 (1996).

\bibitem{MT} M.\ Matsumoto and T.\ Nishimura,
ACM Trans. on Mod. and Comp. Sim., {\bf 8} (1998) 3.

\bibitem{Panneton}
F.\ Panneton, P.\ L'Ecuyer, M.\ Matsumoto,
ACM Trans. Mathem. Software {\bf 32(1)} (2006) 1.

\bibitem{LFSR113}
P.\ L'Ecuyer, Math. of Comp., {\bf 68} (1999) 261.

\bibitem{TestU01Paper}
P.\ L'Ecuyer, R.\ Simard,
ACM Trans. Mathem. Software, {\bf 33(4)} (2007) article 22.

\bibitem{Grothe}
H.\ Grothe, Statistical Papers, {\bf 28} (1987) 233.

\bibitem{Lecuyer90}
P.\ L'ecuyer, Comm. of the ACM, {\bf 33(10)} (1990) 85.

\bibitem{NiederrBook}
H.~Niederreiter, in {\it Monte Carlo and Quasi-Monte Carlo Methods in
Scientific Computing}, ed. H. Niederreiter and P. J.-S. Shiue,
Lecture Notes in Statistics, Vol. 106 (Springer-Verlag, 1995).

\bibitem{AlgSite} 
\verb#http://www.comphys.ru/barash/rng_sse2.zip#

\bibitem{CLN}
\verb#http://www.ginac.de/CLN/#

\bibitem{TestU01}
P.\ L'Ecuyer, R.\ Simard, {\it TestU01: A Software Library in ANSI C for
Empirical Testing of Random Number Generators (2002)}, Software
user's guide, \verb#http://www.iro.umontreal.ca/~simardr/testu01#
\verb#/tu01.html#

\bibitem{CombinedLCG}
P.\ L'Ecuyer, Oper. Res., {\bf 47} (1999) 159.

\bibitem{LEcuyerMain}
P.\ L'Ecuyer,
Chapter 4 of the Handbook of Simulation, Jerry Banks Ed., Wiley, 1998,
pp. 93--137.

\bibitem{MCerrors}
A.M.\ Ferrenberg, D.P.\ Landau, Y.J.\ Wong,
Phys.\ Rev.\ Lett. {\bf 69} (1992) {3382};
L.N.\ Shchur,
Comp.\ Phys.\ Comm. {\bf 121-122} {(1999)} {83};
{P.\ Grassberger},
{Phys. Lett.} {\bf 181} {(1993)} {43};
{L.N.\ Shchur, J.R.\ Heringa, H.W.J.\ Bl\"ote},
{Physica A} {\bf 241} {(1997)} {579};
{L.N.\ Shchur, H.W.J.\ Bl\"ote},
{Phys.\ Rev.\ E} {\bf 55} {(1997)} {R4905};
{F.\ Schmid, N.B.\ Wilding},
{Int.J.Mod.Phys. C} {\bf 6} {(1995)} {781}.

\end{thebibliography}
\end{document}